\ifdefined\WORDCOUNT
\documentclass[twocolumn,showpacs,preprintnumbers,amsmath,amssymb,prl,nofootinbib]{revtex4-1}
\else
\documentclass[twocolumn,showpacs,preprintnumbers,amsmath,amssymb,prl]{revtex4-1}
\fi

\usepackage{amssymb}
\usepackage{epsfig,amsmath,graphics}
\usepackage{epstopdf}
\usepackage{graphicx}

\newcommand{\Bbias}{B_0}

\newcommand{\objectSubscript}{o}
\newcommand{\lensSubscript}{\textsc{\tiny{$\ell$}}}
\newcommand{\imageSubscript}{i}

\newcommand{\refocusSubscript}{\text{min}}

\newcommand{\tLensDuration}{\delta t}

\newcommand{\tObjectTime}{t_\objectSubscript}
\newcommand{\tImageTime}{t_\imageSubscript}

\newcommand{\temperature}{T}

\newcommand{\dvObject}{\Delta v_\objectSubscript}
\newcommand{\dvLensed}{\Delta v_\lensSubscript}
\newcommand{\dvLensedEstimate}{(\Delta v_\lensSubscript)_\text{bound}}

\newcommand{\temperatureEstimate}{\temperature_\text{bound}}

\newcommand{\dxObject}{\Delta x_\objectSubscript}
\newcommand{\dxLensed}{\Delta x_\lensSubscript}
\newcommand{\dxImage}{\Delta x_\imageSubscript}

\newcommand{\aberrationHeating}{\delta\mkern-1mu A}
\newcommand{\sizeRatio}{\gamma}

\ifdefined\WORDCOUNT
    \newcommand{\HideEq}[1]{}
\else
    \newcommand{\HideEq}[1]{#1}
\fi

\def\be{\begin{equation}}
\def\ee{\end{equation}}

\begin{document}

\title{Matter wave lensing to picokelvin temperatures}

\def\stanfordAffiliation{Department of Physics, Stanford University, Stanford, California 94305}

\author{Tim Kovachy}
\author{Jason M. Hogan}
\author{Alex Sugarbaker}
\author{Susannah M. Dickerson}
\author{Christine A. Donnelly}
\author{Chris Overstreet}
\author{Mark A. Kasevich}
\email{kasevich@stanford.edu}
\affiliation{\stanfordAffiliation}

\date{\today}

\begin{abstract}
Using a matter wave lens and a long time-of-flight, we cool an ensemble of $^{87}$Rb atoms in two dimensions to an effective temperature of less than $50^{+50}_{-30}$~pK. A short pulse of red-detuned light generates an optical dipole force that collimates the ensemble. We also report a three-dimensional magnetic lens that substantially reduces the chemical potential of evaporatively cooled ensembles with high atom number. By observing such low temperatures, we set limits on proposed modifications to quantum mechanics in the macroscopic regime.  These cooling techniques yield bright, collimated sources for precision atom interferometry.
\end{abstract}


\ifdefined\WORDCOUNT
\else
   \maketitle
\fi

The observation of low-temperature phenomena has historically enabled new discoveries \cite{Osheroff1996, Chu1997, Ketterle2001}.  Accordingly, significant experimental effort has been dedicated to reaching increasingly cold temperatures.  In this work, we report the demonstration of a cooling protocol to prepare ensembles of $^{87}$Rb atoms with effective temperatures of tens of pK, which is to our knowledge the lowest kinetic temperature ever measured.  We present a new approach to atomic thermometry that allows us to resolve such low temperatures.

Atomic thermometry with kinetic temperatures in the pK range tests quantum mechanics at macroscopic scales \cite{Nimmrichter2013, Bassi2013}.  Our results place bounds on proposed modifications to quantum mechanics that predict the breakdown of quantum superpositions in the macroscopic regime \cite{Nimmrichter2013}.

Additionally, the ability to reach lower temperatures has driven numerous advances in precision measurement \cite{Cronin2009, Dickerson2013, Wilpers2002}, quantum information \cite{Mandel2003}, and quantum simulation \cite{Bakr2009}.  Our realization of a cooling protocol to achieve effective temperatures of tens of pK meets a critical need for a new generation of atomic sensors with dramatically increased sensitivity \cite{Hogan2011}.  These sensors are expected to have a broad scientific reach, with applications including gravitational wave detection \cite{Hogan2011, Graham2013}, tests of general relativity \cite{Dimopoulos2008, Muller2008, Muntinga2013}, and precision geodesy \cite{Hogan2011}.

Evaporative cooling offers one route to low kinetic temperatures \cite{Leanhardt2003}. An alternative cooling method \footnote{Here `cooling' is understood to mean kinetic temperature reduction.}, often called delta-kick cooling, is to freely expand an atom cloud and then reduce its velocity spread with a collimating lens \cite{Chu1986, Ammann1997, Marechal1999, Morinaga1999, Myrskog2000, Meystre2001, Cornell1991, Smith2008}.  Compared to evaporation, lensing typically requires less time and avoids intrinsic atom loss, but does not increase phase space density. The lens is implemented by a transient harmonic potential, realized magnetically \cite{Monroe1990, Marechal1999,Shvarchuck2002}, electrostatically \cite{Kalnins2005}, or optically \cite{Chu1986}.  In previous work, lensing has yielded effective temperatures as low as ${\sim} 1~\text{nK}$ \cite{Muntinga2013, McDonald2013}.

In this work, we use a sequence of lenses to continuously manipulate the RMS velocity of ensembles of $^{87}$Rb atoms through a minimum value of $\Delta v <70~\text{$\mu$m/s}$, corresponding to effective temperature $\temperature=~m\Delta v^2/k_{\text{B}}<~50~\text{pK}$ for atomic mass $m$ \cite{Metcalf1999}.  The dipole lensing potential \cite{Chu1986, Metcalf1999} is generated from the transverse intensity profile of a vertically-propagating Gaussian beam, providing cooling in two dimensions.

This cooling performance is facilitated by several advances.  We use a long expansion time {$>$ 1 s} before the application of the dipole lens, which greatly improves its cooling capability.  Additionally, we realize a cooling protocol that minimizes the influence of imperfections of the lensing potential on the ensemble temperature.  For instance, we implement a dual-stage sequence in which a magnetic lens provides initial cooling, reducing the heating from aberrations in the second-stage dipole lens.

The potential cooling performance of the dipole lens depends on the available expansion time. Consider an initial atom ensemble (condensate or thermal state) with RMS size $\dxObject$ and velocity spread $\dvObject$, allowed to expand for an object time $\tObjectTime$ before application of the lens potential. After the lens is applied, the RMS velocity is $\dvLensed$ and the temperature ratio is $\eta\equiv(\dvLensed/\dvObject)^2$.  For an ideal harmonic potential that has been tuned to minimize $\dvLensed$ (the collimation condition), $\eta$ is bounded by $\eta_c = (\dxObject/\dxLensed)^2\equiv \sizeRatio^2$, where $\dxLensed$ is the RMS size of the ensemble when the lens is applied and $\sizeRatio$ is the size ratio \cite{Chu1986}. Correlations between position and velocity in the initial ensemble (e.g, arising from mean field interactions during expansion) can lead to temperatures that are lower than this bound \cite{virtualSourceNote}. To achieve low temperatures, it is beneficial to have a long expansion time so that $\dxLensed\approx\dvObject \tObjectTime \gg \dxObject$.

An ideal harmonic lens (frequency $\omega$) exerts a force $F_\textsc{\tiny $H$}=-m\omega^2 x$, where $x$ is the transverse position. For the dipole potential lens, the lens duration $\tLensDuration$ is short (delta-kick limit, $\omega \tLensDuration \ll 1$), so we may approximate its effect as an impulse that changes the atom's velocity by $\delta v(x) =-\omega^2 \tLensDuration \, x$.  The lens focal time is defined by $1/f\equiv \omega^2 \tLensDuration$ so that a point source of atoms expanding for time $f$ would be perfectly collimated.

In order to measure these very low temperatures, we use a new method of atomic thermometry.  At pK temperatures, the time necessary for the ensemble size to noticeably increase can be very long ($>10~\text{s}$), making time-of-flight expansion an ineffective probe of temperature. To circumvent this, we extend the duration of the dipole-potential lens interaction beyond the collimation condition to refocus the ensemble.  As in optics, the minimum achievable image size after refocusing is a measure of the degree of collimation.  Thus, we can infer the collimated temperature of the atom ensemble from the refocused cloud size.  An analogous method has been used to measure the temperature of electron beams \cite{Engelen2013}.

To formalize this relationship, we solve the quantum Liouville equation for the evolution of an arbitrary initial state during the lensing sequence. In the delta-kick limit, this reduces to solving the classical Liouville equation \cite{supplemental, Dubetsky2006, Giese2014}.  To account for aberration in the lens, we assume a general lens force $F(x)$. We find that the minimum refocused size $(\dxImage)_\text{min}$ sets a bound on the minimum velocity spread $\dvLensed$ achievable at collimation. By this metric, the minimum velocity variance for the lens (including aberrations) can be inferred by:
\HideEq{\be
\label{eqn:temperatureBound}
\dvLensedEstimate^2\equiv\frac{(\dxImage)_\text{min}^2}{\tImageTime^2}=\dvLensed^2+\aberrationHeating \gtrsim \dvLensed^2
\ee}
\noindent where $\tImageTime$ is the time between the lens and detection (`image time'), and $\aberrationHeating$ arises from lens aberrations present during refocusing.  For a wide class of aberrations (including those encountered in this work), $\aberrationHeating$ is positive, so $\dvLensedEstimate^2$ provides an upper bound on the collimated temperature \footnote{Correlations between lens aberrations and the initial atom distribution can yield $\aberrationHeating < 0$, but even in the worst case the correction to $\dvLensedEstimate$ is within our stated uncertainty.}.

\begin{figure}
\begin{center}
\includegraphics[width=\columnwidth]{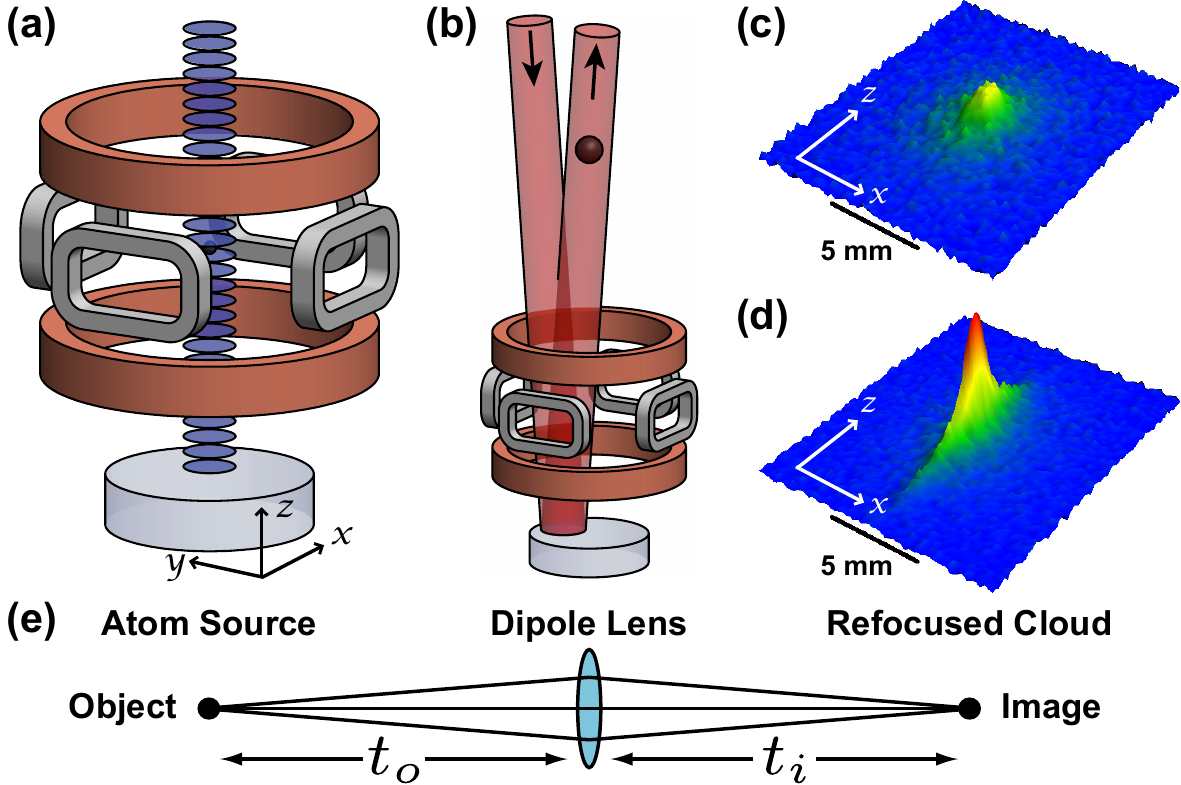}
\caption{ \label{Fig:apparatus} (a) Schematic of the apparatus (including vertically-oriented quadrupole trap, horizontal TOP coil pairs, and blue-detuned launching lattice).  (b) A $3~\text{W}$ laser, $1/e^{2}$ radial waist $\sigma=3.4~\text{mm}$, $1.0~\text{THz}$ red-detuned from the $^{87}$Rb $D_2$ line, acts on the atom cloud as a dipole lens (the $\sim$$1~\text{mrad}$ beam angle is  exaggerated for clarity).  (c) Fluorescence image of a $1.6~\text{nK}$ cloud after $2.8~\text{s}$ of free-fall.  (d) The distribution in (c) refocused using the dipole lens.  There is no observed axial heating.  (e) Optical analogy showing the object, lens, and image, with object distance $\tObjectTime$ and image distance $\tImageTime$.}
\end{center}
\end{figure}

\begin{figure}
\begin{center}
\includegraphics[width=\columnwidth]{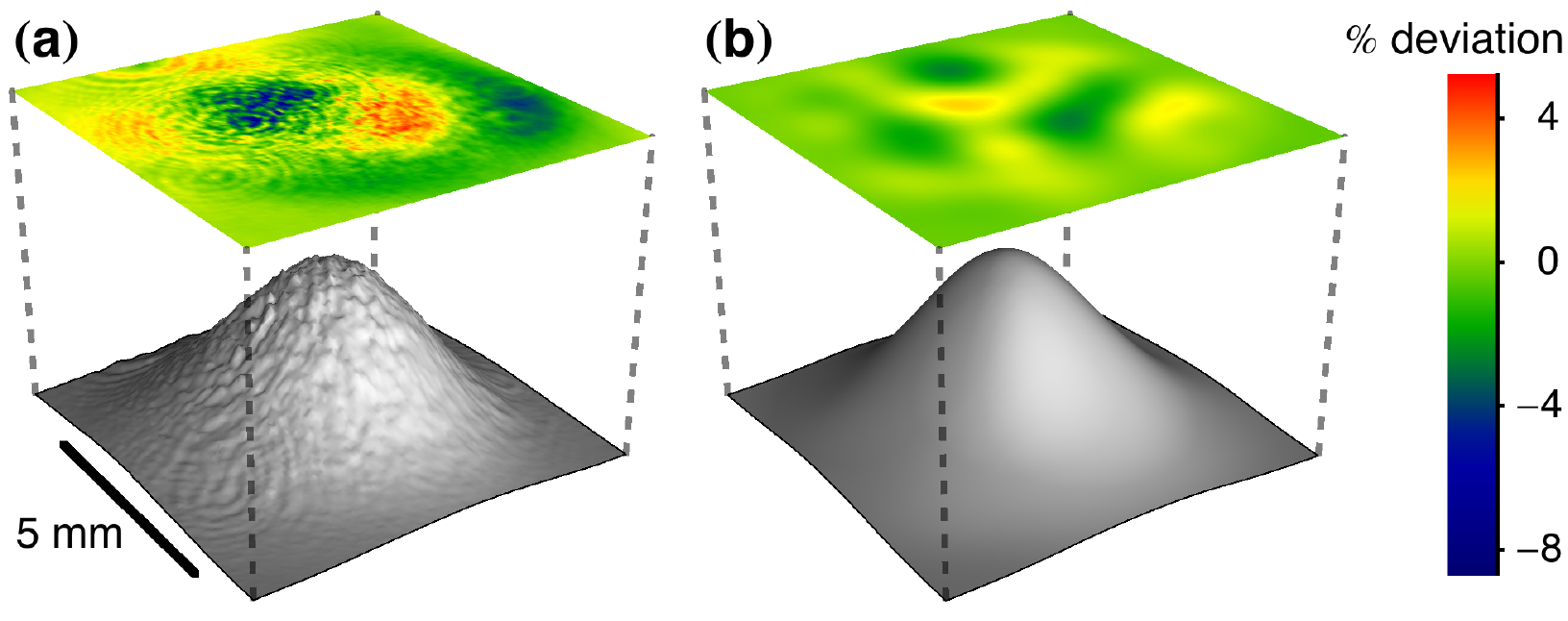}
\caption{ \label{Fig:BeamAberration} Comparison of the dipole lens beam intensity profile after numerical paraxial wave propagation of the measured profile by (a) $0.25~\text{m}$ and (b) $16.25~\text{m}$. The residuals of fitting a 2D Gaussian are shown above each beam profile.  The beam is initially spatially filtered by propagation through an optical fiber.}
\end{center}
\end{figure}

The cooling performance demonstrated here depends critically on an optics configuration that reduces spatial intensity perturbations on the dipole lensing beam.  Perturbations with spatial frequency $\kappa$ produce forces $\propto \kappa$, so high spatial frequency perturbations {($\kappa\sigma > 1$ for radial waist $\sigma$)} are particularly detrimental \cite{supplemental}.  For example, for our beam waist $\sigma=3.4~\text{mm}$, a $1\%$ perturbation with $\kappa\sim (100~\mu\text{m})^{-1}$ can result in a spurious force comparable in magnitude to the lensing force, substantially heating the cloud. To avoid this, the beam propagates for $16~\text{m}$ or more from the collimation lens (retroreflected after $10.6~\text{m}$) before interacting with the atoms [Fig.~\ref{Fig:apparatus}(b)], allowing high spatial frequencies to diffract from the beam (Fig.~\ref{Fig:BeamAberration}). With $\tLensDuration = 30~\text{ms}$ and $\tObjectTime = 1.1~\text{s}$, the lens substantially refocuses the atoms at a time $\tImageTime = 1.8~\text{s}$ later [Fig.~\ref{Fig:apparatus}(d)].

The atom source is a cloud of $10^5$ $^{87}$Rb atoms with initial RMS size $\dxObject = 56~\mu\text{m}$ \footnote{Determined by Gaussian fits to absorption images.}  and an effective temperature of $1.6 \pm 0.1~\text{nK}$ [Fig.~\ref{Fig:apparatus}(c)].   To prepare this ultracold source, we evaporate in a time-orbiting potential (TOP) trap [Fig.~\ref{Fig:apparatus}(a)].  The atoms are further cooled with a magnetic lens (details follow) and prepared in a magnetically insensitive state.  We then launch them upwards into a $10~\text{m}$ vacuum tube with a chirped optical lattice \cite{Dickerson2013, latticeNote}.  After $2.8~\text{s}$ \footnote{The release from the magnetic trap occurs $100~\text{ms}$ before the end of the lattice launch.}, the atoms fall back down, and we image them with a vertical fluorescence beam onto two CCD cameras (the $y$-axis camera images the $x$-$z$ plane, and $x$-axis camera images the $y$-$z$ plane).

To evaluate the performance of the optical lens, we vary the lens duration and measure the width of the lensed cloud.  As the lens acts only transversely, we bin the corresponding images in the vertical dimension and analyze in 1D.  Extracting cloud widths requires accounting for the point spread function (PSF) of the imaging system. We fit all imaged clouds to a Gaussian profile convolved with a smooth representation of the PSF \cite{supplemental}.

\begin{figure}
\begin{center}
\includegraphics[width= \columnwidth]{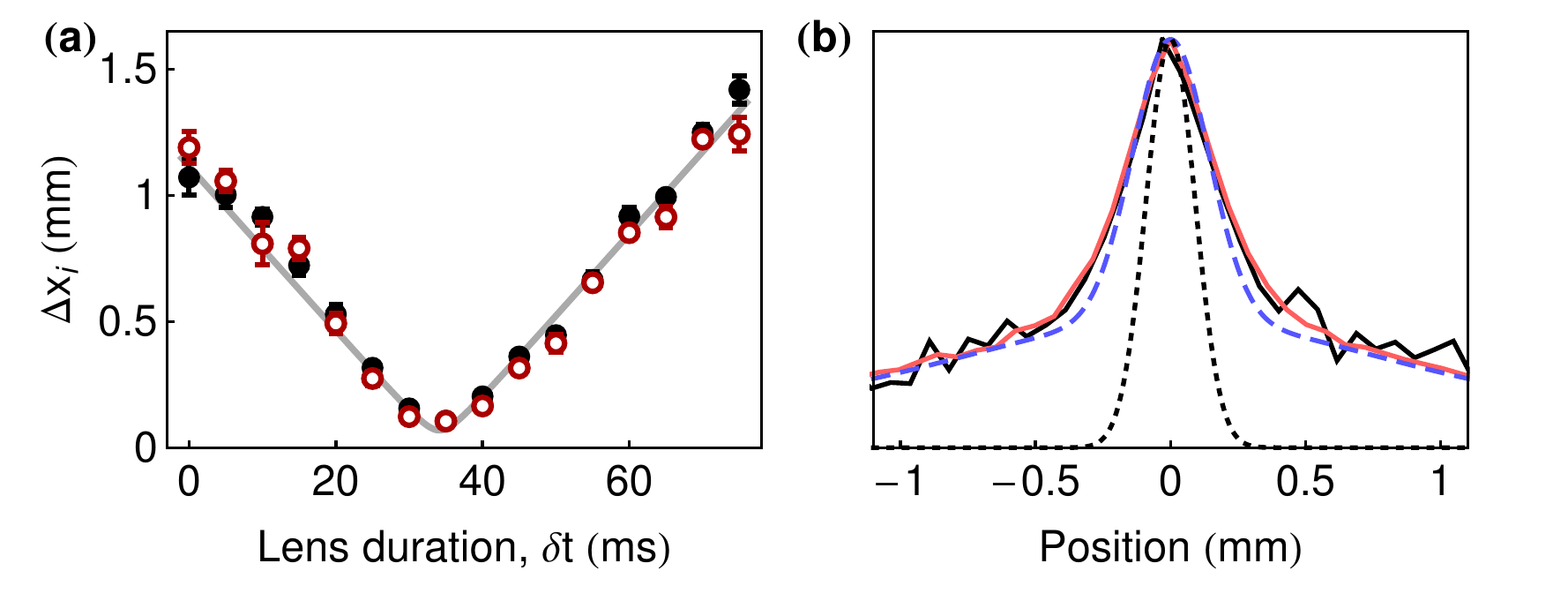}
\caption{ \label{Fig:CloudSizeVLensDuration}
(a) Filled black (open red) points denote measured RMS cloud widths on the $x$-axis ($y$-axis) camera. Each point is the weighted mean of Gaussian fits to $6$ experimental shots. The dashed gray curve is a simultaneous fit to the measurements from both cameras and reports a minimum size of $70 \, \mu$m at a lens duration of $34 \,$ms.  (b)  Vertically-binned images comparing the transverse size of a $90 \pm 10~\mu$m cloud used to characterize the PSF (solid red/gray) to a cloud refocused $2.8$~s later (solid black). The good overlap indicates high-fidelity refocusing. Dotted black: Gaussian profile extracted from a fit of the refocused cloud. The fit accounts for the broadening and distorting effects of the PSF (dashed blue).}
\end{center}
\end{figure}

To characterize the PSF, we fit a cloud with a known, small size; this fixes the PSF parameters for subsequent analysis. We use a cloud imaged after a short drift time ($100~\text{ms}$; the time needed to reach the fluorescence imaging region) as the small source [Fig.~\ref{Fig:CloudSizeVLensDuration}(b)]. To directly measure this cloud's size, we image it with a low-aberration imaging system \footnote{This imaging system was not used for primary data collection because of its comparatively poor photon collection efficiency.}.  The measured width of $90 \pm 10~\mu$m is consistent with an extrapolation from the known cloud parameters at the end of the TOP sequence.

Figure \ref{Fig:CloudSizeVLensDuration}(a) shows the fitted transverse cloud size $\dxImage$ versus lens duration $\tLensDuration$ for the two camera axes, demonstrating the continuous variation of the atom cloud through collimation and refocus.  For this data, the lens is applied $\tImageTime = 1.8~\text{s}$ before detection.  We fit the data with the predicted cloud size \cite{supplemental},
\HideEq{\be
\dxImage^2 = (\dxImage)_\text{min}^2 + \tfrac{1}{m^2}\Delta F^2\tImageTime^2 \left(\tLensDuration - \tLensDuration_\refocusSubscript\right)^2
\label{Eq:CloudSizeVsLensDuration}
\ee}
\noindent where $\Delta F^2$ is a fitting parameter characterizing the variance of the lensing force (including any aberrations), and $\tLensDuration_\refocusSubscript$ is the lens duration to refocus the cloud.  Even for the smallest refocused cloud size, the chemical potential is negligibly small ($\sim 0.2~\text{pK}$), so chemical potential does not limit our ability to refocus.  In fact, since the cloud expands vertically during the drift time, chemical potential would not prevent the cloud from being refocused to smaller than its initial transverse size.

The point at $\tLensDuration=35~\text{ms}$ is nearest to the fitted refocusing time and sets the best bound on the achievable collimation temperature $\temperature$. From Eq.~\ref{eqn:temperatureBound}, we find that ${\dvLensedEstimate \equiv (\dxImage)_\text{min}/\tImageTime= 65 \pm 20~\mu\text{m/s}}$ for the $x$-axis and $70 \pm 25~\mu\text{m/s}$ for the $y$-axis. These bound the effective temperature at collimation to below $\temperatureEstimate \equiv m \dvLensedEstimate^2/k_{\text{B}} = 40^{+40}_{-20}~\text{pK}$ and $50^{+50}_{-30}~\text{pK}$ for the $x$ and $y$ axes, respectively.  This $\temperatureEstimate$ estimate includes extra heating $\aberrationHeating$ that arises between collimation time $\tLensDuration_c$ and refocus $\tLensDuration_\text{min}$. Since heating from aberrations scales as $\tLensDuration^2$, we can estimate $\temperature$ by multiplying the aberration contribution $(\dvLensedEstimate^2-\sizeRatio^2\dvObject^2)$ by $(\tLensDuration_c/\tLensDuration_\text{min})^2 \approx (0.6)^2$, yielding effective temperatures of $30^{+10}_{-10}~\text{pK}$ and $35^{+15}_{-10}~\text{pK}$ for the $x$ and $y$ axes, respectively \cite{supplemental}.  The temperature uncertainties result primarily from the standard deviation of the measured cloud sizes, likely caused by shot-to-shot fluctuations in the strength of the lens (e.g., due to fluctuations in optical power or alignment). Uncertainties in the measured PSF do not contribute significantly.

These kinetic temperature measurements are a sensitive probe of quantum mechanics at the macroscale.  For a wide class of proposed modifications to quantum mechanics, the mechanism that leads to decoherence of macroscopic quantum superpositions also causes a free gas to undergo a small amount of spontaneous heating \cite{Nimmrichter2013, Bassi2013}.  For example, in the theory discussed in \cite{Nimmrichter2013}, the heating results from spontaneous momentum kicks that also lead to spontaneous wavefunction localization.  These theories can therefore be bounded using precise heating rate measurements of ensembles of ultracold atoms.  In our experiment, spontaneous heating would lead to diffusion of the atom cloud during the long drift time, limiting our ability to refocus the cloud.  The measured refocused cloud size $(\dxImage)_\text{min}$ constrains the heating rate for $^{87}$Rb to $20 \pm 30~\text{pK/s}$ \cite{supplemental}.

Our ability to transversely cool an atom cloud to low effective temperatures and to refocus the cloud after long drift times has many applications \cite{Muntinga2013, Dickerson2013}, including terrestrial atom interferometers with interrogation times previously thought to require microgravity.  Refocusing the atom cloud to its original size allows us to relaunch the ensemble, extending the effective free-fall time to $5.1~\text{s}$.  One or more of these relaunches could be integrated into an atom interferometer, possibly leading to $> 10~\text{s}$ interrogation times on Earth. Similarly, a series of relay lenses (or an initial collimation lens) could be integrated with light-pulse atom interferometry to maintain a small transverse cloud size at the beamsplitter pulses, even for very long interrogation times.  This would ensure a homogeneous atom optics beam intensity across the cloud, which is critical for large momentum transfer atom interferometry \cite{Chiow2011}.

\begin{figure}
\begin{center}
\includegraphics[width=\columnwidth]{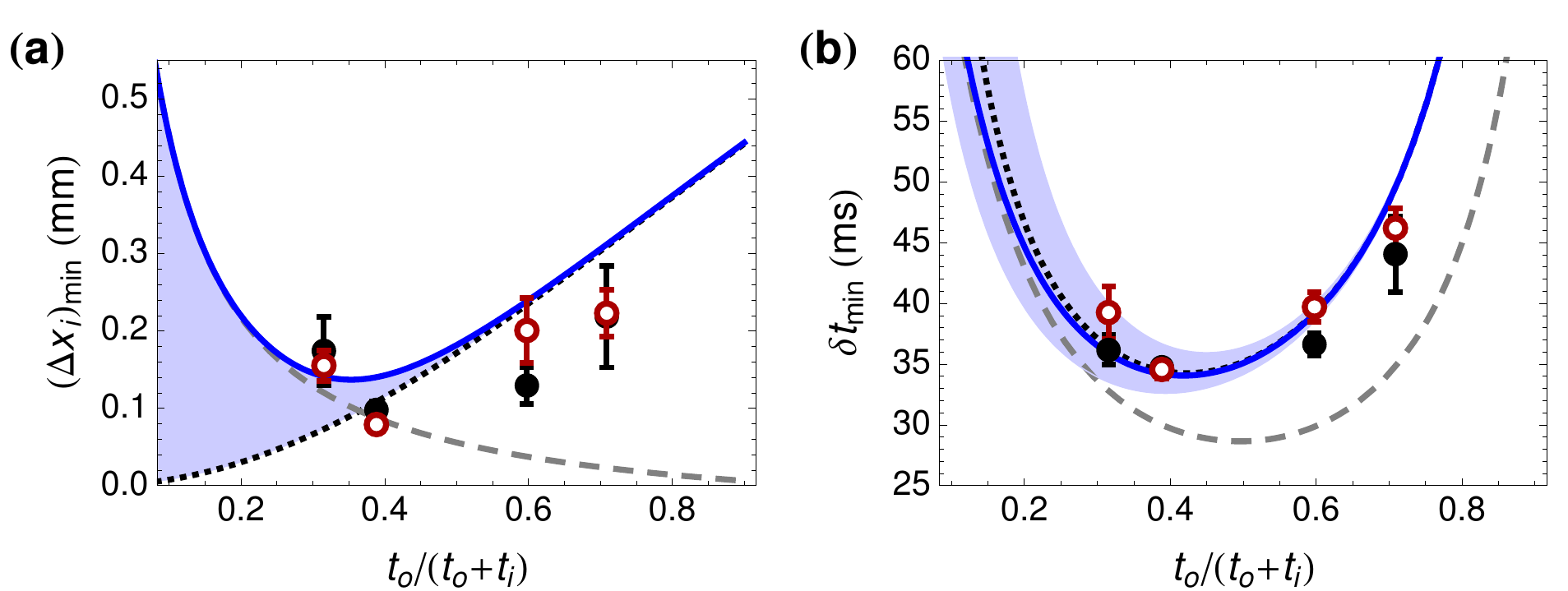}
\caption{ \label{Fig:LensDurationToRefocus} (a) Minimum RMS width of the cloud as a function of the fractional object time. (b) The lens duration required to refocus the atom cloud ($\tLensDuration_\refocusSubscript$) as a function of the fractional object time. The filled black (open red) points represent measurements on the $x$-axis ($y$-axis) camera. Solid blue: expected behavior for a cloud of finite initial size ($56 \, \mu \text{m}$) and no initial $x$-$v$ correlations in an optical beam with a Gaussian profile. The blue shaded regions represent the corresponding ranges possible with correlations. The curve in (a) has no free parameters. For (b), the optical power is a free parameter that fits to $2.8~\text{W}$.  Dashed gray: expected behavior for an ideal harmonic potential of the same strength as at the center of the beam \cite{supplemental}. Dotted black: expected behavior for a cloud with zero initial size subject to Gaussian aberrations.}
\end{center}
\end{figure}

To characterize imperfections in the dipole lensing potential and the corresponding deviations from ideal lens behavior, we measure the refocused cloud size $(\dxImage)_\text{min}$ and corresponding lens duration $\tLensDuration_\refocusSubscript$ for various object times $\tObjectTime$, with the total atom drift time held constant (Fig.~\ref{Fig:LensDurationToRefocus}). Each point is the result of a fit of Eq.~\ref{Eq:CloudSizeVsLensDuration} to a scan of the lens duration [like Fig.~\ref{Fig:CloudSizeVLensDuration}(a)] at one of four fractional object times: $\tObjectTime/(\tObjectTime+\tImageTime) = 0.32$, $0.39$, $0.60$, and $0.71$. Also shown is the ideal harmonic lens scaling for $(\dxImage)_\text{min}$ and $\tLensDuration_\refocusSubscript$. Neglecting $x$-$v$ correlations, the focal time $f_\refocusSubscript \equiv (\omega^2\tLensDuration_\refocusSubscript)^{-1}$ satisfies the thin lens formula from geometric optics $\tfrac{1}{f_\refocusSubscript} = \tfrac{1}{\tImageTime} + \tfrac{1-\sizeRatio^2}{\tObjectTime}$ (the $\sizeRatio$ correction results from finite velocity spread and vanishes in the point source limit $\sizeRatio\ll 1$ \footnote{An identical correction arises in optics with a finite source divergence (e.g., a Gaussian laser beam).}) and the image size $(\dxImage)_\text{min} = \dxObject \tfrac{\tImageTime}{\tObjectTime}\sqrt{1-\sizeRatio^2}$ scales as the magnification of the lens $\tfrac{\tImageTime}{\tObjectTime}$.

The deviation of the data in Fig.~\ref{Fig:LensDurationToRefocus} from the harmonic lens theory results primarily from large-scale aberrations due to the Gaussian profile of the optical potential. Modeling the lens potential as a 2D Gaussian, we calculate $(\dxImage)_\text{min}$ and $\tLensDuration_\refocusSubscript$ assuming Gaussian initial ensemble velocity and position distributions (Fig.~\ref{Fig:LensDurationToRefocus}) \cite{supplemental}.  Although the cooling performance of the lens is partially limited by the finite expansion time $\tObjectTime$, further extending $\tObjectTime$ would not improve cooling performance, since a larger $\dxLensed$ would increase the effect of Gaussian aberration.

To reach the lowest temperatures, it is necessary to minimize the impact of anharmonicities of the dipole lensing beam.  To do this, we implement a dual-stage lensing sequence, pre-cooling the atoms with an initial magnetic lens.  This increases the effective $f/\#$ of the dipole lens by reducing the duration $\tLensDuration$ required for collimation.  It has the added benefit of cooling along the third axis not addressed by the dipole lens.

The magnetic lens is performed by abruptly releasing tightly-confined atoms into a shallow harmonic TOP trap potential [Fig.~\ref{Fig:apparatus}(a)] \cite{Minogin1998}.  Subsequently turning off the shallow trap when the ensemble has reached its maximum size yields a colder cloud \cite{Monroe1990}. Synchronizing the radial ($\rho$) and vertical ($z$) oscillations to optimize 3D cooling requires a trap frequency ratio of $\omega_z / \omega_\rho = \left(n_z + 1/2\right) / \left(n_\rho + 1/2\right)$ for integers $n_z$ and $n_\rho$ (we use $n_z = 3$, $n_\rho = 2$).  In the absence of gravity, the irrational ratio $\omega_z / \omega_\rho = 2 \sqrt{2}$ of the TOP trap makes perfect synchronization impossible, but with gravity the ratio is tunable by selecting the appropriate quadrupole gradient \cite{Hodby2000}.

\begin{figure}
\begin{center}
\includegraphics[width=\columnwidth]{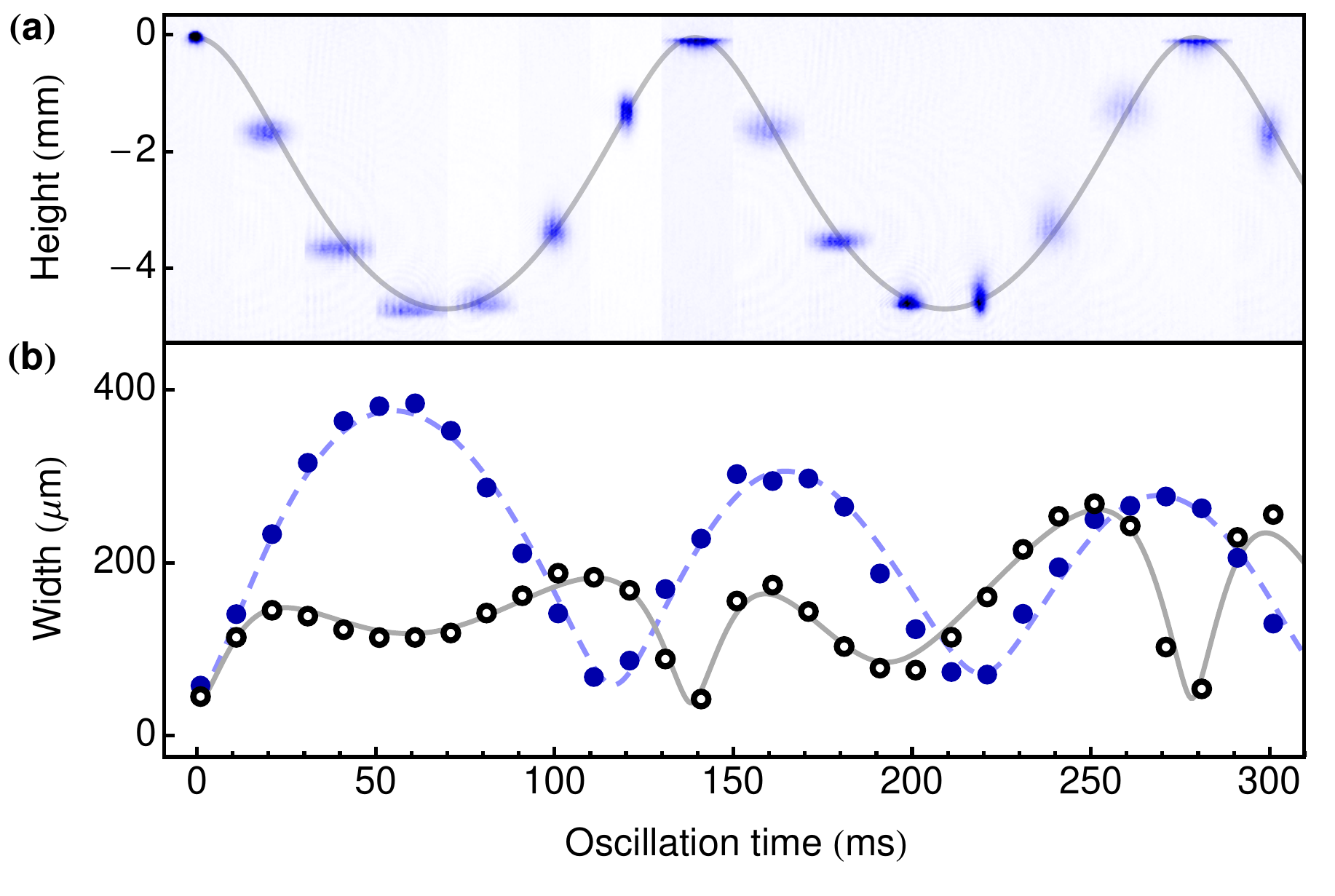}
\caption{ \label{Fig:LatticelessTOPOscillations} Magnetic lensing in the TOP trap. (a) Absorption images of the ensemble oscillating in the trap. (b) Radial (filled blue circles) and vertical (open black circles) RMS cloud widths. Theory curves are based on numerical solutions for trajectories of non-interacting particles in the exact TOP potential. The solid grey curves are simultaneous fits to the center-of-mass trajectory and the vertical width, with free parameters for the TOP potential (radial quadrupole gradient $\nabla B$, spinning bias field $\Bbias$, and vertical position) as well as initial first and second moments of the vertical distribution. The dashed blue curve results from a 2D Monte Carlo simulation of the atom distribution using the fitted parameters (including $\nabla B = 20.9 \pm 0.1~\text{G}/\text{cm}$, $\Bbias = 6.9 \pm 0.3~\text{G}$) but no free parameters. The initial radial size is scaled from the fitted vertical size by the ratio of the measured initial cloud widths.}
\end{center}
\end{figure}

Figure~\ref{Fig:LatticelessTOPOscillations} shows the evolution of a cloud while in the lens (widths are from 2D Gaussian fits). The center of mass oscillates vertically because the atoms start above the minimum of the shallow trap. The initial evaporated source used here has a release temperature of $1.4~\mu\text{K}$ (dominated by chemical potential).  The optimized lens duration of $162~\text{ms}$ gives a minimum effective temperature of $(T_\rho, T_z) = (50, 40)~\text{nK}$ and reduced chemical potential. The $1.6~\text{nK}$ source used as input to the dipole lens is derived from an identical magnetic lens sequence, but with a deeper initial evaporation cut.

Our combined magnetic/dipole lensing sequence has the potential to reach even colder temperatures.   Fundamentally, the diffraction-limited collimation temperature for a wavepacket with size $\dxLensed$ at the lens is determined by the minimum velocity width allowed by the uncertainty principle, in this work $\sim$10 fK (for the $\sim 400~\mu$m clouds).  Future work will seek to achieve these limits.

\ifdefined\WORDCOUNT
\else
\begin{acknowledgments}
The authors would like to thank Sheng-wey~Chiow, Naceur~Gaaloul, and Jan~Rudolph for valuable discussions and contributions to the apparatus. TK acknowledges support from the Fannie and John Hertz Foundation; TK, AS, and CAD from the NSF GRFP; and TK and CO from the Stanford Graduate Fellowship. This work was supported in part by NASA GSFC Grant No. NNX11AM31A.
\end{acknowledgments}
\fi

\bibliographystyle{apsrev4-1}

\ifdefined\WORDCOUNT
    \end{document}
\else
\fi

\bibliography{LensingarXiv.bbl}

\begin{thebibliography}{44}
\bibitem [{lat()}]{latticeNote}%
  \BibitemOpen
  \href@noop {} {}\bibinfo {note} {The lattice's depth is 40 times the recoil
  energy and its $1/e^{2}$ radial waist is 1.5~mm.}\BibitemShut {Stop}%
\end{thebibliography}

\end{document}